# Spectroscopic realization of large surface gap in a doped magnetic topological insulator

Xiao-Ming Ma[1]*, Yufei Zhao[1]*, Ke Zhang[2]*, Rui'e Lu[1], Jiayu Li[1], Qiushi Yao[1], Jifeng Shao[1], Xuefeng Wu[1], Meng Zeng[1], Yu-Jie Hao[1], Shiv Kumar[3], Zhanyang Hao[1], Yuan Wang[1], Xiang-Rui Liu[1], Huiwen Shen[1], Hongyi Sun[1], Jiawei Mei[1], Koji Miyamoto[3], Taichi Okuda[3], Masashi Arita[3], Eike F. Schwier[3], Kenya Shimada[3], Ke Deng[1], Cai Liu[1], Yue Zhao[1], Chaoyu Chen[1]†, Qihang Liu[1,4]†, Chang Liu[1]†

[1]*Shenzhen Institute for Quantum Science and Engineering (SIQSE) and Department of Physics, Southern University of Science and Technology, Shenzhen 518055, China*

[2]*Department of Physical Science, Graduate School of Science, Hiroshima University, Higashihiroshima, Hiroshima 739-0046, Japan*

[3]*Hiroshima Synchrotron Radiation Center, Hiroshima University, Higashihiroshima, Hiroshima 739-0046, Japan*

[4]*Guangdong Provincial Key Laboratory for Computational Science and Material Design, Southern University of Science and Technology, Shenzhen 518055, China*

*These authors contributed equally to this work.
†Corresponding author.
Electronic mails: chency@sustech.edu.cn; liuqh@sustech.edu.cn; liuc@sustech.edu.cn

**Realization of the quantum anomalous Hall effect and axion electrodynamics in topological materials are among the paradigmatic phenomena in condensed matter physics[1-4]. Recently, signatures of both phases are observed to exist in thin films of MnBi$_2$Te$_4$, a stoichiometric antiferromagnetic topological insulator[5-8]. Direct evidence of the bulk topological magnetoelectric response in an axion insulator requires an energy gap at its topological surface state (TSS)[3]. However, independent spectroscopic experiments[9-12] revealed that such a surface gap is absent, or much smaller than previously thought, in MnBi$_2$Te$_4$. Here, we utilize angle resolved photoemission spectroscopy and density functional theory calculations to demonstrate that a sizable TSS gap unexpectedly exists in Sb-doped MnBi$_2$Te$_4$. This gap is found to be topologically nontrivial, insensitive to the bulk antiferromagnetic-paramagnetic transition, while enlarges along with increasing Sb concentration. Our work shows that Mn(Bi$_{1-x}$Sb$_x$)$_2$Te$_4$ is a potential platform to observe the key features of the high-temperature axion insulator state, such as the topological magnetoelectric responses and half-integer quantum Hall effects.**

Magnetic topological insulators (MTIs) are condensed matter systems possessing long-range magnetic order but remains topologically nontrivial[13-15]. Compared to nonmagnetic TIs whose topological surface states (TSSs) manifest a gapless Dirac cone that is protected by time reversal symmetry ($T$), the TSSs of MTIs could open an energy gap if out-of-plane ferromagnetic (FM) order exists at the surface[16]. The presence of this gap in MTIs is of central importance to reveal the bulk topological magnetoelectric response, as each of these gapped surfaces hosts an anomalous Hall conductivity (AHC) that is quantized to a half of $e^2/h$. The addition or subtraction of such half-quantized AHCs at the top and the bottom surface of a two-dimensional slab gives rise to two macroscopic transport phenomena in two dimensions, namely, the QAH effect, and the zero Hall plateau[17]. Compared with the QAH insulator where the topological nature is accessed by its chiral edge states[1], the experimental signature of axion insulator phase lies in an intermediate phase to two QAH states in which the AHCs at the gapped top and bottom surfaces cancel each other, yielding zero Hall conductivity. Albeit much less explored than the QAH state, axion insulators shed light on the fundamental understanding of topological insulators as bulk magnetoelectrics[14], and is potentially practical even in the astronomical search of the dark axions, a quasiparticle candidate of the long-sought nonbaryonic dark matter.[18]

The newly discovered Van der Waals magnetic compounds $(MnBi_2Te_4)(Bi_2Te_3)_n$ ($n$ = 0, 1, 2, …) are thus far the only stoichiometric material system that enables both the QAH and the axion insulator states. The ground-state magnetic orders of the $n$ < 2 compounds are found to be $A$-type antiferromagnetic (AFM), with out-of-plane moments coming from the central Mn planes of the septuple-layer (SL) building blocks[5,19-25]. Along with strong band inversion, these compounds are predicted to be three-dimensional AFM TIs and strong candidates of axion insulators with TSS gaps at their natural cleaving planes[8,16,26]. Intriguingly, angle-resolved photoemission spectroscopy (ARPES) and scanning tunneling spectroscopy (STM) measurements uncovered diminished or vanished surface state gaps in single crystals of the undoped "parent" compounds[9-12,27-31], raising the question on whether the impurities, defects, and possible surface structural and magnetic reconstruction plays a role in realizing the macroscopic quantum phases[9]. Above all, it is still controversial if a surface state gap exists in any of the Mn-Bi-Te family at all[9-12,27-35].

Antimony is a convenient choice of nonmagnetic atomic dopant in the Bi-based topological materials. In $Bi_2(Se, Te)_3$, Sb effectively introduces holes to the otherwise $n$ doped system and drives the system towards the topologically trivial side continuously, while maintaining the Dirac cone before the topological phase transition because the $T$-symmetry is always preserved[36]. Here, we demonstrate via systematic ARPES measurements that the situation in $MnBi_2Te_4$ is fundamentally different. A sizeable, global surface state gap opens in Sb-doped $MnBi_2Te_4$ single crystals. A mere 10% of Sb dopants was able to raise the gap size up to 100 meV. This gap keeps immune from the AFM ground state to the high temperature paramagnetic (PM) state, while increases monotonically with the dosage of Sb. With assistance of density functional theory (DFT) calculations, we identified the signature of electronic structure during the magnetic phase transition, and confirmed the nontrivial topological nature within the doping range studied. Possible origins of the anomalous surface gap are also discussed.

First of all, we distinguish the bulk and surface dispersion and demonstrate the existence of the surface state gap in Mn(Bi$_{1-x}$Sb$_x$)$_2$Te$_4$ via systematic ARPES measurements. Fig. 1 shows our ARPES data taken on a typical sample with a nominal $x$ value of 0.075 at $T = 10$ K (below $T_N$). Figs. 1**a-c** present the raw and second derivative ARPES $k$-$E$ maps along the $\bar{\Gamma}$-$\bar{M}$ high symmetry direction taken under three representative photon energies, corresponding to two consecutive bulk Γ points and a bulk Z point in between. Great care was taken in these measurements to ensure that the $\bar{\Gamma}$ point is missed along the $\bar{\Gamma}$-$\bar{K}$ direction by at most 0.2° (< 0.008 Å$^{-1}$) for all photon energies. A clear, sizable energy gap is observed at the crossing point of the otherwise linear bands in three different photon energies with identical value, in drastic contrast to the case of undoped MnBi$_2$Te$_4$ where the gap is diminishing at the Dirac point. From Fig. 1**d** we found that the gap is a global one, which opens under all photon energies measured, covering more than two out-of-plane Brillouin zones (Fig. 1**e**). There are five visible bands (defined in Fig. 1**b**) near the apparent gap. The bulk nature of the BV band is proven via its strong and periodic $k_z$ dispersive behavior seen in Figs. 1**d-e**. The SV and SC bands, on the other hand, exhibit no discernable dispersion across a $k_z$ range of ~5π/$c$, endorsing their surface origin. We can therefore unambiguously conclude that the surface state of Mn(Bi$_{1-x}$Sb$_x$)$_2$Te$_4$ is gapped.

In Fig. 2 we examine quantitatively the temperature evolution of the bulk and surface bands from the ARPES data on a prototypical Sb-doped sample (Sample S2, $x_{norminal}$ = 0.05, Figs. 2**a-d**), as well as the theoretical AFM and PM electronic structures of bulk Mn(Bi$_{1-x}$Sb$_x$)$_2$Te$_4$ at $x_{cal}$ = 0.056 (1/18) calculated by DFT (Figs. 2**e-h**). The results of other samples with different Sb concentrations are shown in Fig. S3. To establish a direct comparison with the spectral function of ARPES, we unfold the band structure from supercell approach[37] to simulate the disorder effects of Sb doping and the local moments in the PM phase. As shown in Figs. 2**e-h**, while the slightly doped Mn(Bi$_{1-x}$Sb$_x$)$_2$Te$_4$ for the AFM phase does not change much, the long-wavevector spectral density away from $E_F$ for the PM phase looks fuzzy, informing the extent to which the translational symmetry is retained. The most profound difference of the band dispersion between AFM and PM phases near $E_F$ occurs at the Z point, where the BC1 and BC2 bands merge into a single BC band at the high-temperature PM phase. Therefore, we choose to perform ARPES measurements at a photon energy of 6.36 eV, focusing on the bulk $Z_2$ point ($k_z \sim 5\pi/c$). Indeed, we see in Figs. 2**a-d** that the BC1 and BC2 bands come closer to each other as temperature rises, merge into a single BC band right at $T_N$, and finally keep a constant binding energy for $T > T_N$. Therefore, the band evolution through the AFM-PM magnetic phase transition is unambiguously observed.

Surprisingly, the surface gap, on the other hand, remains essentially unchanged for all temperatures measured, across $T_N$ from 16 K to 35 K (Fig. 2**d**). In Fig. S4 we graph the temperature evolution of this gap for two other samples with different $x$, reproducing again a constant-sized gap up to 150 K. These observations reveal an unexpected fact that the surface state gap of Mn(Bi$_{1-x}$Sb$_x$)$_2$Te$_4$ is insensitive to the change of temperature, regardless of its bulk magnetic phase. Another important feature about the gap is that it increases in samples with higher Sb concentration. This behavior is elaborated in Fig. 3 where the sizes of both the TSS and the bulk gap are compared at $T > T_N$ for five samples with different Sb dosage, corresponding to $0 \leq x_{norminal} \leq 0.1$. Since the actual carrier concentration varies greatly even within the same growth batch (see Section S1 in the Supplementary Information), the doping levels are calibrated using the binding energies at the center of the TSS gap ($E_c$) instead of $x_{norminal}$. This procedure is justified by the knowledge that Sb atoms are effective $p$ dopants of the system, pushing the Fermi level downward in a rigid band shifting scenario except for the gap region. Samples S1 to S5 are named and ordered by this means. From Figs. 3**a-e**, we see that the gap size increases as Sb doping increases, from ~0 meV at Sample S1 ($x = 0$, $E_c = -272$ meV) to ~100 meV at Sample S5 ($E_c = -156.5$ meV). Moreover, the size of the bulk gap between the BV and the BC bands is also increasing with doping, from 192 meV at Sample S1 to 265 meV at Sample S5. Fig. 3**f** further shows that these gap sizes evolve almost linearly with $E_c$.

So far we have demonstrated the presence and the anomalous temperature-doping dependence of a global energy gap in the TSS of Mn(Bi$_{1-x}$Sb$_x$)$_2$Te$_4$. This gap is nearly constant across a wide temperature range, insensitive to the bulk magnetic transition, but enlarges monotonically with increasing Sb concentration. Before looking into the possible origins of these gapped phases, we first point out that the system remains topologically nontrivial within the doping range studied. Experimentally, we notice that the doping evolution of both the bulk and the TSS gap is rather smooth, no sudden change of gap size is observed (Fig. 3**f**). This hints at a common topological nature between the undoped and doped samples. Theoretically, our DFT calculations[37,38] show that the topological phase transition (TPT) in Mn(Bi$_{1-x}$Sb$_x$)$_2$Te$_4$ occurs at a much larger $x$. Fig. 3**g** marks the TPT by displaying the calculated band gap evolution at Γ and Z[37]. The growing gap at Z with increasing $x$ is consistent with our ARPES results, while the gap at Γ closes at $x_c \approx 0.75$, where the system undergoes a transition to the normal-insulator state. Similar calculation results using the virtual crystal approximation (VCA)[38] are shown in the Supplementary Information, in which $x_c$ is found to be about 0.65. Therefore, we believe that for $0 < x < 0.1$, the system stays in the topological regime.

We note that the TSS of the $x = 0$ parent compound is gapless, or at least has a gap that is much smaller than DFT prediction[9-11]. Although a comprehensive explanation on the intact Dirac cone of MnBi$_2$Te$_4$ is not reached, these results point to the possibility that the structural and/or magnetic structure on the surface of the system is fundamentally different from that in the bulk[9]. Thus, it is natural to speculate that Sb doping in Mn(Bi$_{1-x}$Sb$_x$)$_2$Te$_4$ suppresses the surface spin re-orientation and somehow restore the surface FM, which is supported by the ferrimagnetic state recently found by transport measurements in Sb doped samples[39]. However, this argument is not consistent with our ARPES results in the PM phase, because a magnetic gap arising from this mechanism is supposed to vanish at $T > T_N$. Similarly, previous work proposed that the Dirac electronic states could couple to the conduction electrons through the Ruderman-Kittel-Kasuya-Yosida (RKKY) exchange interaction among the PM impurities and thus induce weak ferromagnetism[40]. However, it is uncertain that such FM state above $T_N$ could quantitively cause a ~100 meV gap as found, which inevitably leads to a significantly higher transition temperature at the surface. The residual FM at the surface is supposed to be captured by monitoring the $s_z$ components of the TSS from the spin-ARPES measurements[41]. We performed such measurements on an undoped sample and found large, antiparallel $z$ polarization of the spin at $\bar{\Gamma}$. As shown in Fig. S5, the $s_z$ components are observed to be as large as 30%, in drastic contrast to the case of nonmagnetic TIs where $s_z = 0$ at $\bar{\Gamma}$. However, due to the limitation of the energy resolution of spin-ARPES, one cannot rule out the bulk contribution when the penetration effect counts the Mn(Bi$_{1-x}$Sb$_x$)$_2$Te$_4$ layer close to the surface predominately[42].

Another possible scenario is that this gap does not result from the surface FM, but rather the effect of dephasing effect and Coulomb scattering from charged impurities, such as vacancies (see Fig. 4). We note that previous studies on several kinds of *doped* topological insulators showed a Dirac gap in which the charged impurities are not negligible, no matter whether the dopant is magnetic or not[43-45]. Now we examine this possibility by constructing a model Hamiltonian and calculate directly the spectral function of the TSS. In the absence of long-range magnetic order, the gapless TSS is depicted by the effective Hamiltonian $H(\mathbf{k}) = \hbar v_F(\mathbf{k} \times \boldsymbol{\sigma})_z$ with Fermi velocity $v_F$ and spin Pauli matrices $\boldsymbol{\sigma}$. Considering the combined scattering effects of charge impurities and background-induced quasiparticle dephasing, we obtain the surface Green's function $G_{\pm}^r(E, \mathbf{k}) = [E \pm \hbar v_F \mathbf{k} - iIm\Sigma^r(\mathbf{k})]^{-1}$ with the self-energy[46]

$$-Im\Sigma^r(\mathbf{k}) = \frac{2\alpha^2 n_i}{\mathbf{k}^2}\left[\hbar v_F|\mathbf{k}| + \frac{2\alpha^2 \hbar}{\pi}\left(\frac{1}{\tau_n} + \frac{1}{\tau_s}\right)\right],$$

where $\alpha = e^2/4\pi\varepsilon_0 \hbar v_F$ is the "fine-structure constant" of the TSS, $n_i$ is the charged impurity concentration proportional to the concentration of Sb dopants, and $\tau_{n,s}$ are the normal and spin dephasing time, respectively. The dephasing effects[46,47] contain the normal part $\tau_n$ caused by electron-electron and electron-phonon interactions, and the spin part $\tau_s$ by random distribution of Mn local magnetic moments. By calculating the spectral function of TSS $A(E, \mathbf{k}) = -\sum_{\pm} Im G_{\pm}^r(E, \mathbf{k})/\pi$, we find a ~30 meV gap-like feature with a reasonable concentration of $n_i = 5 \times 10^{10}$ cm$^{-2}$ (Fig. 4c). We note that such speculation simultaneously fulfills the two anomalous features of the TSS gap, i.e., remaining above *T$_N$* and increasing upon doping, but it does not perfectly fit the observed shape of the gapped bands. More evidence that exposes the underlying mechanism of the anomalous gap is thus called for.

In summary, a sizeable global energy gap in the topological surface state of Sb-doped magnetic topological insulator Mn(Bi$_{1-x}$Sb$_x$)$_2$Te$_4$ (0 < *x* < 0.1) is discovered experimentally. Our systematic ARPES measurements found that this gap increases in size to more than 100 meV as $x$ increases, but remains constant in both the low-temperature AFM and the high-temperature PM phases. The transition between the two phases is identified by the merging of two bulk conduction bands at $T_N$, observed both in our ARPES measurements and DFT calculations. Restoration of surface FM, weak ferromagnetism introduced by RKKY interaction, as well as the combined effect of charged impurity and quasiparticle dephasing are discussed as possible origins of the TSS gap. Remarkably, these possibilities do not hinder the topological nontriviality of the gap. It is nonetheless a large, robust surface state gap with inverted band order in the bulk. Taken collectively, we suggest that Mn(Bi$_{1-x}$Sb$_x$)$_2$Te$_4$, comprising a single massive Dirac cone caused either by the broken $T$-symmetry or the charged impurities, might be thus far the simplest material system to observe the signatures of the high-temperature axion insulator state, such as the topological magnetoelectric effects and half quantization quantum Hall effects.

## Acknowledgements

We thank Prof. Xin-Cheng Xie, Haiwen Liu, Hai-Zhou Lu, Dr. Zhi Wang and Dr. Wen Huang for inspiring discussions. ARPES experiments were performed with the approval of the Hiroshima Synchrotron Radiation Center (HSRC), Hiroshima, Japan under Proposals 19BG044, 19BU002, 19BU005 and 19BU012. Work at SUSTech was supported by the National Natural Science Foundation of China (NSFC) (No. 11504159, No. 11674149, and No. 11874195), NSFC Guangdong (No. 2016A030313650), the Guangdong Innovative and Entrepreneurial Research Team Program (No. 2016ZT06D348 and No. 2017ZT07C062), the Guangdong Provincial Key Laboratory of Computational Science and Material Design (Grant No. 2019B030301001), the Shenzhen Key Laboratory (Grant No. ZDSYS20170303165926217), the Technology and Innovation Commission of Shenzhen Municipality (Grants No. JCYJ20150630145302240 and No. KYTDPT20181011104202253), and Center for Computational Science and Engineering of SUSTech. J. S. is supported by the SUSTech Presidential Postdoctoral Fellowship. C. C. is supported by the Shenzhen High-level Special Fund (No. G02206304, G02206404). Ch. L. acknowledges support from the Highlight Project (No. PHYS-HL-2020-1) of the College of Science, SUSTech.

**Author Contributions**

Ch. L. conceived the project; Ch. L., Q. L., and C. C. supervised the project. X.-M. M., Y.-J. H., R. L., M. Z., and Ch. L. grew the samples. X.-M. M., K. Z., R. L., M. Z., Y.-J. H., S. K., Z. H., Y. W., X.-R. L., H. Shen, K. M., T. O., M. A., E. F. S., K. S., K. D., Cai L., C. C., and Ch. L. carried out the ARPES experiments. R. L., J. S., Yue Z., and Ch. L. performed the transport and magnetic measurements. X. W. and Yue Z. did the STM measurements. Yufei Z., Q. Y., H. Sun and Q. L. carried out the DFT calculations; J. L. and Q. L. carried out the dephasing model calculations. Ch. L., Q. L., C. C., X.-M. M., Yufei Z. and J. L. wrote the paper with input from all authors. All authors commented on the paper.

*Methods*

**Sample growth**

The Mn(Bi$_{1-x}$Sb$_x$)$_2$Te$_4$ single crystals were grown by the conventional high-temperature solution method[12,48]. The Mn (purity 99.98%), Bi (purity 99.999%), Sb(99.99%) and Te (99.999%) blocks were weighed with a molar ratio of Mn: Bi: Sb: Te = 1: 11.3(1-$x$): 11.3$x$: 18, grounded in an agate mortar and then placed in an alumina crucible. The alumina crucible was sealed in a quartz tube under the argon environment. The assembly was first heated up in a box furnace to 950 °C, held for 10 hrs, then subsequently cooled down to 700 °C in 10 hrs and further cooled down slowly to 575 °C in 100 hrs. After this heating procedure, the quartz tube was taken out quickly and then decanted into the centrifuge to remove the excess flux from the single crystals.

**Transport and magnetic measurements**

The structure of the crystals was checked by X-ray diffraction with Cu K$\alpha$ radiation at room temperature using a Rigaku Miniex diffractometer. The crystal structures of various doped single crystals are consistent with that of pure MnBi$_2$Te$_4$, but the 2$\theta$ change to high values, demonstrating the successful doping. Resistivity measurements were performed by a Quantum Design (QD) Physical Properties Measurement System (PPMS) with a standard six-probe method, with a drive current of 8 mA. The current flows in the *ab* plane and the magnetic field is perpendicular to the current direction. Linear electrical contacts were made by Au wires using Ag epoxy as a glue. Magnetic measurements were performed using the QD PPMS with the Vibrating Sample Mangetometer (VSM) mode. Temperature dependent magnetization results were collected with an external magnetic field of 0.1 T, both along and perpendicular to the [0001] direction of the sample.

**ARPES measurements**

The ARPES data in this article was collected at the Hiroshima Synchrotron Radiation Center (HSRC) of Japan. The laser ARPES measurements were performed at HSRC with a VG Scienta R4000 electron analyzer and a laser photon source of 6.36 eV. The energy and angular resolution were better than 3 meV and less than 0.05°. $k_z$ variation measurements (Fig. 1) were performed at Beamline 9A of HSRC with a VG Scienta R4000 electron analyzer with linearly polarized lights from 6.5 eV to 23 eV at a temperature of 10 K (below $T_N$).. The energy and angular resolutions were set at 15-30 meV and 0.2°, respectively. Spin-resolved ARPES data (Fig. S4) were taken at Beamline 9B of HSRC[49]. The samples were measured at 20 K (below $T_N$). The spin polarization was detected by the high-efficiency very low energy electron diffraction (VLEED) spin detectors using preoxidized Fe(001)-$p$(1×1)-O targets. The energy and angular resolutions were set at 80 meV and 1°, respectively. All the samples were cleaved *in-situ* along the (0001) crystal plane in an ultrahigh vacuum better than $7 \times 10^{-11}$ mbar.

**First-principles calculations**

First-principles calculations are carried out with Vienna Ab-initio Simulation Package (VASP)[50], with the PBE generalized gradient approximation (GGA+U)[51,52] exchange-correlation functional. For a direct comparison with ARPES results, we take the effective band structure (EBS) method[37] to transfer the complex dispersion in a supercell BZ into the spectrum density in a primitive BZ. The lattice parameters of Mn(Bi$_{1-x}$Sb$_x$)$_2$Te$_4$ supercells are taken as a linear interpolation of the MnBi$_2$Te$_4$ and MnSb$_2$Te$_4$ experimental lattice parameters according to the doping concentration $x$, and the internal coordinates of the atoms are fully relaxed. We do not relax the cell shape and volume because the GGA gives an inadequate estimation of the lattice constants for the coupling of van der Waals between two SLs, especially the one in the $z$ direction. Spin-orbit coupling (SOC) is included self-consistently and the $U$ parameter is selected to be 5 eV for Mn 3$d$ orbitals.

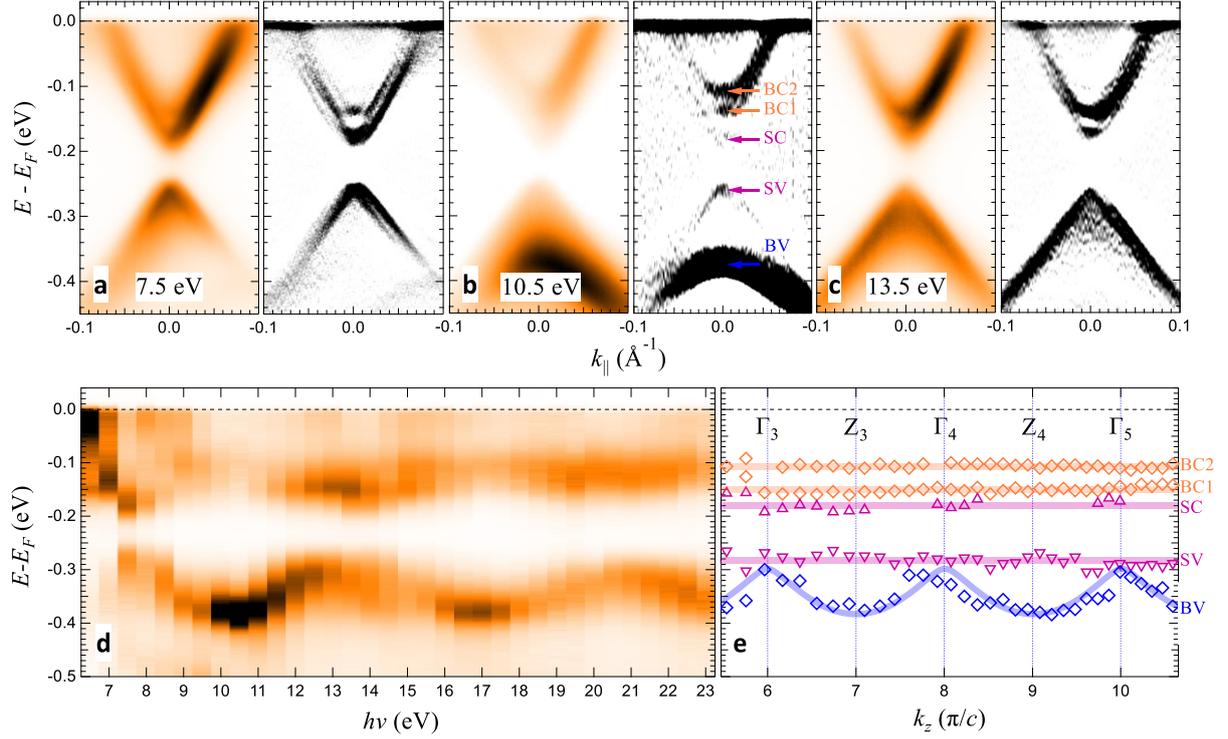

**Fig. 1 Presence of the surface state gap in Mn(Bi$_{1-x}$Sb$_x$)$_2$Te$_4$.** Data taken on a $x_{\text{norminal}} = 0.075$ sample at $T = 10$ K. **a-c**, Raw (left) and second derivative (right) ARPES $k$-$E$ maps on three representative photon energies, close to the bulk $\Gamma_3$ (**a**), $Z_3$ (**b**), and $\Gamma_4$ (**c**) points, respectively. BV: bulk valence band; SV/SC: surface valence/conduction band (bottom/top part of the gapped surface state); BC1/BC2: the two bulk conduction bands seen at $T < T_N$. The persistence of the surface state gap and the evolution of BV is seen clearly. **d-e**, $k_z$ dispersion map of the bands at $\bar{\Gamma}$. Photon energy ranges from 6.5 to 23 eV, corresponding to ~5.5 $\pi/c < k_z <$ ~10.6 $\pi/c$. **d**, Raw EDCs at $\bar{\Gamma}$ vs. incident photon energy. **e**, Extracted band locations vs. $k_z$. Colored lines are guides to the eye.

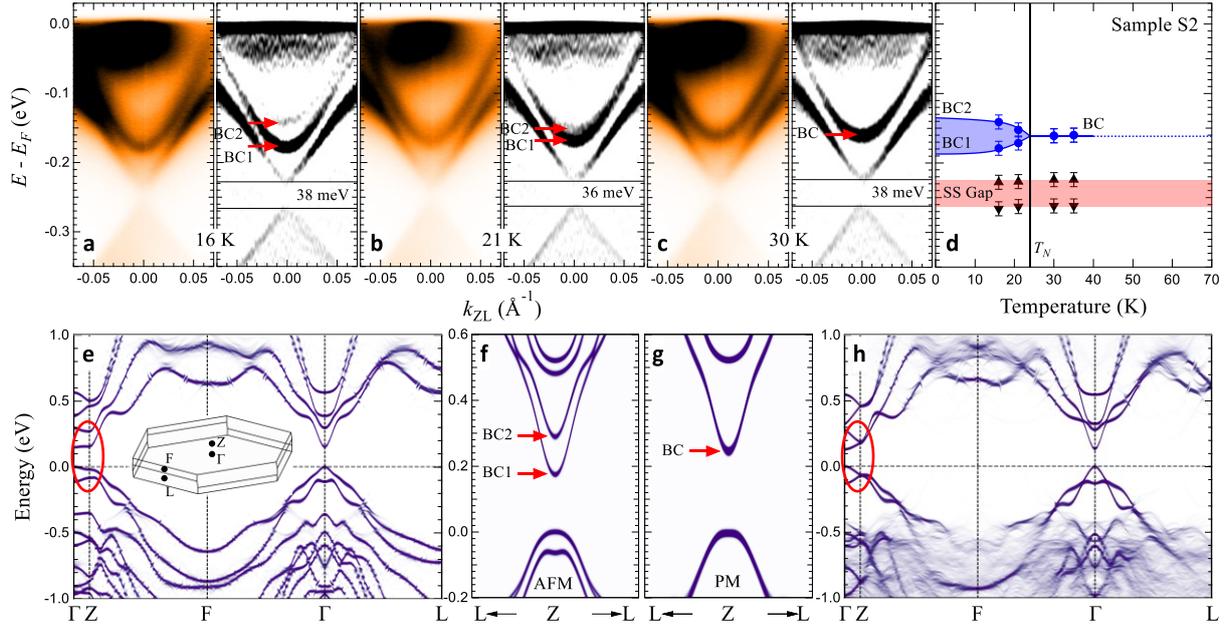

**Fig. 2 Temperature independence of the bulk and surface state gap. a-d**, Temperature evolution of the bands of Sample S2 ($x_{norminal} = 0.05$). Data is taken with a 6.36-eV laser ARPES setup [i.e., at $k_z \sim 5\pi/c$ ($Z_2$)]. **a-c**, Raw (left) and second derivative (right) ARPES $k$-$E$ maps taken at three representative temperatures, below and above the bulk AFM-PM transition temperature $T_N$. **d**, Summary on the temperature evolution of BC1, BC2 and the surface state (SS) gap. It is seen clearly that while BC1 and BC2 merges into a single bulk conduction (BC) band at $T_N$, the SS gap remains essentially unchanged. **e-h**, Effective band structure (EBS) calculation results for the AFM and PM state electronic structure on a $x = 1/18$ (0.056) system. Darker color represents higher spectral weight. **e**, Overall band structure of the AFM state. Inset shows the AFM Brillouin zone with high-symmetry points. Red ellipses in **e** and **h** highlights the difference of bands along Γ-Z. **f**, Focused band structure of the AFM state, graphed along the L-Z-L high-symmetry line close to Z. **g**, Same as **f** but for the PM state. **h**, Same as **e** but for the PM state. The merging of BC1 and BC2 at Z is clearly reproduced.

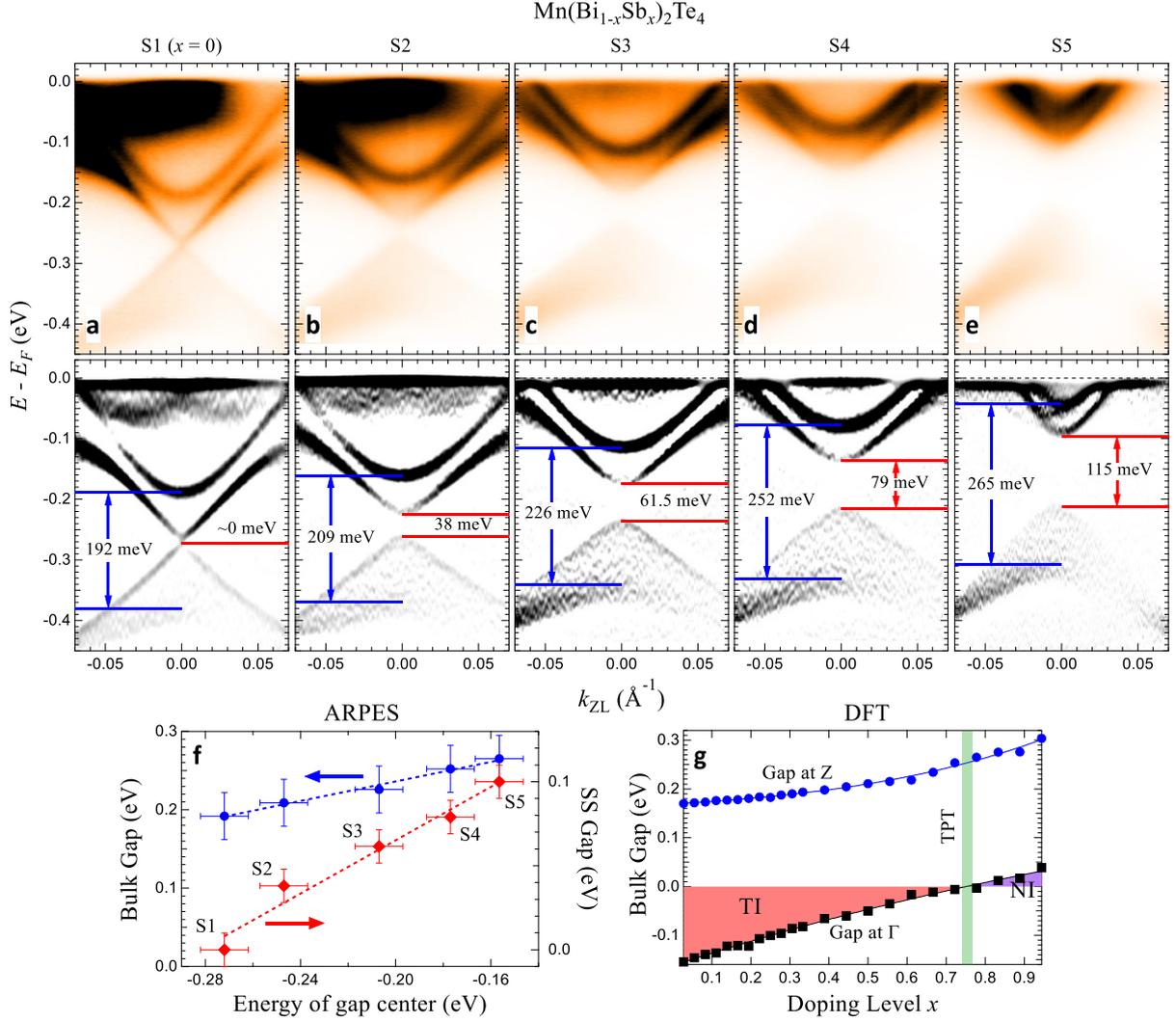

**Fig. 3 Carrier concentration and temperature dependence of the SS gap and the bulk gap.** Data is taken with a 6.36-eV laser ARPES setup at $T$ = 30-35 K (above $T_N$), such that BC becomes a single band. **a-e**, Raw (top) and second derivative (bottom) ARPES $k$-$E$ maps for five samples with different carrier concentrations, ordered by the central energy of the TSS gap (horizontal axis in **g**). Red and blue lines denote the size of the TSS gap and the bulk gap, respectively. **f**, Summary on the carrier concentration dependence of the bulk (blue, left) and the TSS (red, right) gap. Both gaps increase almost linearly with carrier concentration (energy of TSS gap center), albeit at different rates. **g**, Bulk gap size at Γ (black) and Z (blue) calculated via EBS. Possible errors due to different Sb configuration are within symbol size. Solid lines are polynomial fits. TI: topological insulator; NI: normal insulator; TPT: topological phase transition. Determined from the sign of the gap at Γ, Mn(Bi$_{1-x}$Sb$_x$)$_2$Te$_4$ undergoes a TPT from an MTI to a normal magnetic insulator at $x_{c,\text{EBS}} \sim 0.75$. On the other hand, the gap at Z keeps increasing with increasing $x$, consistent with our ARPES result.

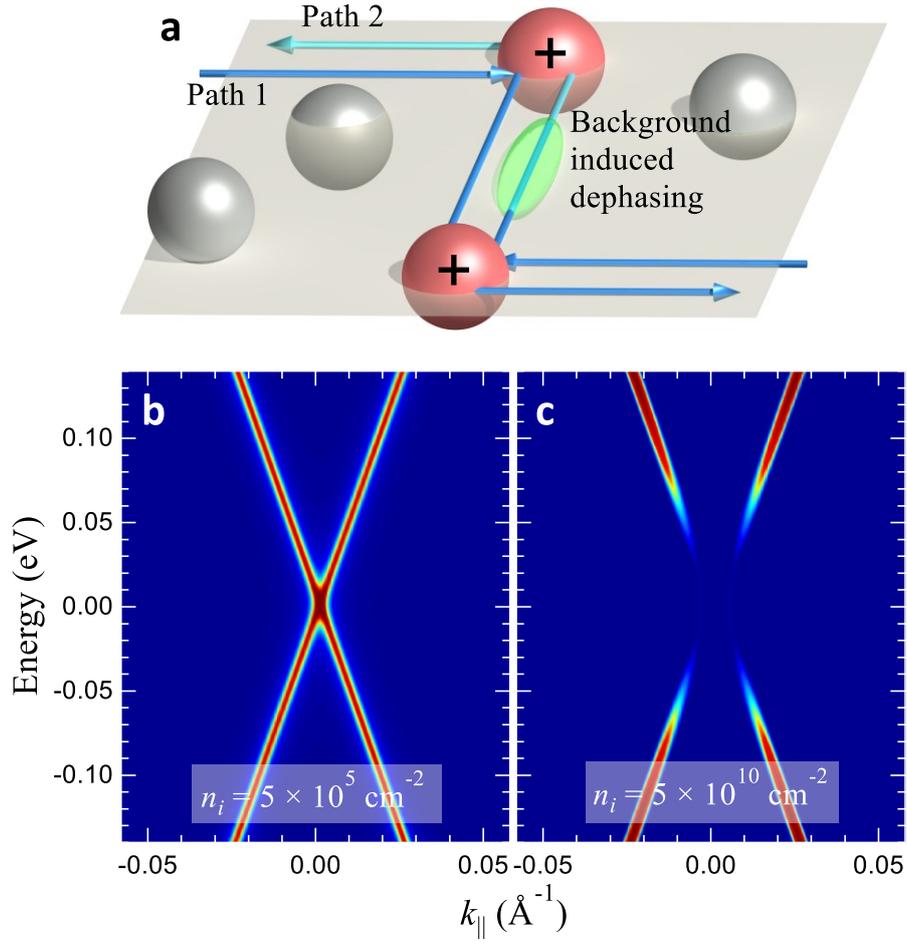

**Fig. 4 Scattering induced gaplike spectrum of the surface state in Mn(Bi$_{1-x}$Sb$_x$)$_2$Te$_4$. a**, Sketch of two time-reversal counterparts of scattering path through two charge impurities ("+"), where a random phase difference produced by fluctuant background emerges in one of the counterparts (Path 2). **b-c**, Color map of the surface spectral function with different impurities concentration, in which parameters are chosen as $v_F = 8.5 \times 10^5$ m/s, $\tau_n = 0.1$ ns, and $\tau_s = 1$ μs.